\documentclass[nofootinbib,notitlepage,superscriptaddress,10pt,aps,pra,twocolumn]{revtex4-1}
\pdfoutput=1

\usepackage[utf8x]{inputenc}
\usepackage{amsmath}
\usepackage{amssymb}
\usepackage{bbm}
\usepackage{graphicx}

\begin{document}

\title{Relative locality in a quantum spacetime and the pregeometry of $\kappa$-Minkowski}

\author{Giovanni AMELINO-CAMELIA}
%\email{Giovanni.Amelino-Camelia@roma1.infn.it}
\affiliation{Dipartimento di Fisica, Universit\`a di Roma ``La Sapienza", P.le A. Moro 2, 00185 Roma, Italy}
\affiliation{INFN, Sez.~Roma1, P.le A. Moro 2, 00185 Roma, Italy}

\author{Valerio ASTUTI}
\affiliation{Dipartimento di Fisica, Universit\`a di Roma ``La Sapienza", P.le A. Moro 2, 00185 Roma, Italy}
\affiliation{INFN, Sez.~Roma1, P.le A. Moro 2, 00185 Roma, Italy}

\author{Giacomo ROSATI}
\affiliation{Dipartimento di Fisica, Universit\`a di Roma ``La Sapienza", P.le A. Moro 2, 00185 Roma, Italy}
\affiliation{INFN, Sez.~Roma1, P.le A. Moro 2, 00185 Roma, Italy}

\begin{abstract}
We develop a new description of the much-studied $\kappa$-Minkowski noncommutative spacetime,
centered on representing on a single Hilbert space not only the $\kappa$-Minkowski coordinates,
but also the associated differential calculus and the $\kappa$-Poincar\'e symmetry generators.
 In this ``pregeometric" representation the relevant operators act on the kinematical Hilbert
 space of the covariant formulation of quantum mechanics, which we argue is the natural framework
 for studying the implications of the step from commuting spacetime coordinates
 to the $\kappa$-Minkowski case, where the spatial coordinates do not commute with the time coordinate.
The empowerment provided by this kinematical-Hilbert space representation allows us
to give a crisp characterization of the ``fuzziness''
of $\kappa$-Minkowski spacetime, whose most striking aspect is a relativity of spacetime locality.
We show that relative locality, which had been previously formulated exclusively in classical-spacetime
setups, for a quantum spacetime takes the shape of a dependence of the  fuzziness
of a spacetime point on the distance at which an observer infers properties of the event that marks the point.
\end{abstract}

\maketitle

\section{Introduction}
It is widely expected that in the quantum-gravity realm, at the interface between
quantum mechanics and general-relativistic
gravity, spacetime should exhibit some new structure characterized by
the Planck length $\ell_P = \sqrt{\frac {\hbar G}{c^3}} \sim 10^{-35}\ \text{m}$.
Among the attempts of formulating this new structure several were based on
the idea of spacetime noncommutativity, and in particular a  ``theory laboratory"
which has been much-studied from this perspective
is $\kappa$-Minkowski spacetime~\cite{majrue,lukieANNALS}, with coordinate
noncommutativity
\begin{equation}
\left[ \hat{x}_j , \hat{x}_0 \right] = i\ell  \hat{x}_j ~ ,~~~~\left[ \hat{x}_j,\hat{x}_k \right] = 0 ~ ,
\label{kappadefCOMMUT}
\end{equation}
which is the main focus of the study we are here reporting.

Our work was mostly inspired by the realization that
after nearly two decades (and hundreds of manuscripts) of research on $\kappa$-Minkowski
the community has still not provided a satisfactory understanding of what
could be implied physically by the noncommutativity (\ref{kappadefCOMMUT}).
Several results were obtained on its indirect implications for the structure of momentum space
and the role played
by the $\kappa$-Poincar\'e Hopf algebra~\cite{majrue,lukieANNALS,lukie1991}
in describing the symmetries of $\kappa$-Minkowski. But for the original objective
of spacetime noncommutativity, the one of providing a characterization of
spacetime fuzziness at the Planck length, the implications of (\ref{kappadefCOMMUT})
remain unclear.

We here report what we feel are significant steps forward in facing
 this long-standing challenge.
 The key ingredient of the strategy of analysis we propose is a novel
 type of ``pregeometric representation" of $\kappa$-Minkowski.
 This idea of pregeometric representation had already been discussed for $\kappa$-Minkowski
 (see, {\it e.g.}, Ref.~\cite{gacmaj}) and was originally conceived also as conceptual tool:
 one could conjecture the emergence of $\kappa$-Minkowski from  quantum gravity
at some  level of effective description, and from this perspective it might be natural
to describe $\kappa$-Minkowski noncommutativity in terms
of a standard Heisenberg quantum mechanics introduced at some deeper level of the description.
One would then seek a relationship between $\kappa$-Minkowski coordinates $\hat{x}_j,\hat{x}_0$ and
the phase space coordinates $\hat{q}_\mu,\hat{\pi}_\mu$ of the pregeometric formulation,
with $\hat{q},\hat{\pi}$ forming standard Heisenberg pairs of conjugate observables on a Hilbert space.
It was already established in previous works that this could be done,
but focusing exclusively on giving such a description of the $\kappa$-Minkowski coordinates.
We here find a new pregeometric description capable of accommodating
not only the $\kappa$-Minkowski coordinates but also
the associated differential calculus and the $\kappa$-Poincar\'e symmetry generators.

As evidence of the empowerment produced by our novel pregeometric description
we exhibit the ability of analyzing for the first time in quantitative
manner the implications of the feature which is most fascinating (and puzzling)
of $\kappa$-Minkowski: the form of (\ref{kappadefCOMMUT}) suggests heuristically
that $\kappa$-Minkowski noncommutativity becomes stronger at larger distances (since $\hat{x}_j$ is on
the right-hand-side of the nontrivial commutation relation),\\
but distances from what?\\
And if it happens to be the distance from the origin of the observer's frame is this
then a preferred-frame picture or somehow still a fully relativistic picture?

Indeed naively (\ref{kappadefCOMMUT}) appears to imply a preferred frame picture,
a picture specialized to a preferred observer in a scenario where relativistic symmetries
(and particularly translational invariance) are broken. But this naive interpretation
is even more puzzling considering the many (though partial) successes of the
description of relativistic transformations in $\kappa$-Minkowski as given
in terms of the $\kappa$-Poincar\'e Hopf algebra.

Within our description these crucial issues can be analyzed without
relying on heuristic/naive reasoning: we can formalize fuzzy points in $\kappa$-Minkowski
 as states on our pregeometric Hilbert
space. And our pregeometric description of
the differential calculus and the  $\kappa$-Poincar\'e generators allows us to describe
relativistic transformations very explicitly, in terms of actions on the pregeometric states.
We thereby obtain conclusive evidence of the fact that $\kappa$-Minkowski
is a fully relativistic spacetime. The feature that was customarily described
naively as ``uncertainties growing with distance" is here properly described
as a novel feature of ``relative locality". Relative locality~\cite{bob,principle}
had been previously observed only in some classical-spacetime setups, and there it
affects locality in the sense that events established to be coincident by nearby observers
may appear to be noncoincident in the description of those events given by distant observers
on the basis of their inferences about the events (such as their observation
of particles originating from the events).
In this first example of relative locality in a quantum spacetime, which we
here provide through our analysis
of $\kappa$-Minkowski, one has a fully relativistic description of how the fuzziness
of events may appear to take different shape depending on the distance from the
events: with a given network of events and a given network of observers one would find
that all observers describe as less fuzzy those events that are near to them
whereas they infer increased fuzzyness for events that are far from them.
All this occurs in a fully relativistic manner, and can be understood as mainly
a manifestation of the peculiarities of translational symmetries
in $\kappa$-Minkowski, which we shall here analyze in detail.

We work for simplicity in a 2D (1+1-dimensional) $\kappa$-Minkowski spacetime,
adopting throughout conventions for the Minkowski metric tensor $g_{\mu\nu} =\{ 1,-1,\}$. And  we adopt units such that the speed-of-light scale (speed of massless particles in the infrared limit) and the Planck constant are $1$ ($c=1$,$\hbar=1$).

\section{Preliminaries on $\kappa$-Minkowski differential calculus and
translation generators}\label{seckappafacts}
We start by summarizing in this section the main facts established in the literature
about the translation transformations
available in the $\kappa$-Minkowski spacetime, adopting the consistent set of conventions
first introduced by Majid and Ruegg in Ref.~\cite{majrue}.
In particular, we describe functions of 2D $\kappa$-Minkowski coordinates
in the time-to-the-right ordered basis of exponentials:
\begin{equation}
f(\hat{x}_1 \hat{x}_0) = \int d^2k\ \tilde{f}(k)e^{ik_1\hat{x}_1}e^{-ik_0\hat{x}_0}~.
\end{equation}
Within this choice of conventions
the generators of $\kappa$-Poincar\'e translations
are conveniently characterized through the following rule of action
on exponentials
\begin{equation}
P_\mu \triangleright e^{ik_1\hat{x}_1}e^{-ik_0\hat{x}_0}= k_\mu e^{ik_1\hat{x}_1}e^{-ik_0\hat{x}_0}~.
\label{traslasingle}
\end{equation}
The fact that the translation sector of the $\kappa$-Poincar\'e Hopf algebra
is itself a Hopf algebra is essentially manifest in the observation that the
rule of action (\ref{traslasingle}) implies that for these translation generators
the action on products of functions is governed by a deformed Leibniz rule
\begin{equation}
\begin{split}
&P_\mu \triangleright f(\hat{x}) g(\hat{x}) \\
& = \left( P_\mu \triangleright f(\hat{x}) \right) g(\hat{x}) \!+\! \left( e^{- \ell \delta_\mu^1 P_0 } \triangleright f(\hat{x}) \right) \left(  P_\mu \triangleright g(\hat{x}) \right)  ~,
\end{split}
\label{coproductPj}
\end{equation}
{\it i.e.} these translation generators have ``non-primitive coproduct"
 $\Delta P_\mu = P_\mu \otimes \mathbbm{1} + e^{-\ell \delta_\mu^1 P_0 }  \otimes P_\mu $.

With this coproduct one can show that the commutator (\ref{kappadefCOMMUT}) is conserved in the
following sense
\begin{equation}
P_\mu \triangleright \left[ \hat{x}_1 , \hat{x}_0 \right] = i \ell P_\mu \triangleright \hat{x}_1\ .
\end{equation}

However, the generators $P_\mu$ are not the only notrivial structure needed for implementing
translation transformations in $\kappa$-Minkowski.
One of course wants the coordinates $\hat{x}_\mu'$ of a translated observer to be linked to the
coordinates of the observer from which the translation is made
by a rule of the type $\hat{x}_\mu' =\hat{x}_\mu - \hat{a}_\mu$,
while enforcing $\kappa$-Minkowski noncommutativity also for the translated
coordinates $\left[ \hat{x}_j' , \hat{x}_0' \right] = i\ell  \hat{x}_j',~\left[ \hat{x}_j',\hat{x}_k' \right] = 0$.
This can be done by describing a the action of translations in
the familiar form
\begin{equation}
T=\mathbbm{1} + {\bf d} \ , \qquad {\bf d}= - i \hat{a}^\ell_\mu P^\mu \ ,
\label{translationGENERATOR}
\end{equation}
but only if the ``translation parameters" have themselves some noncommutativity
properties~\cite{sitarzCALC,majDIFF,oecklDIFF,majidoecklCALC,meljakappa},
which in particular can take the form
\begin{equation}
\left[ \hat{a}_1^\ell , \hat{x}_0 \right] = i\ell \hat{a}_1^\ell ~ ,~~~~\left[ \hat{a}_\mu^\ell , \hat{x}_1\right] = 0 ~ ,\qquad [\hat{a}_0^\ell , \hat{x}_0] = 0 ~.
\label{parametersCOMM}
\end{equation}
One can show that these (\ref{parametersCOMM}) satisfy the conditions for having
a quantum differential calculus, in the sense first introduced by Woronowicz~\cite{woroDIFF}.
And the description of translations based
on (\ref{translationGENERATOR})-(\ref{parametersCOMM}) proved robust
also in work establishing~\cite{kappanoether} the presence of Noether charges
in theories formulated on $\kappa$-Minkowski
with $\kappa$-Poincar\'e symmetries.

\section{A novel pregeometric representation of $\kappa$-Minkowski}\label{pregeomsec}
The notion of pregeometric description which we are here
adopting was proposed in Ref.~\cite{gacmaj}.
Conceptually it can be inspired by the idea (or it can suggest that) spacetime noncommutativity
arises from a more fundamental theory: the more fundamental theory would be needed to analyze
more general quantum-gravity issues but in certain limiting cases (regimes) a description
based solely on spacetime noncommutativity would arise.
Technically a pregeometric description allows to reformulate
the complexity of the $\kappa$-Minkowski commutation relations
in terms of (a few copies of) the familiar Heisenberg algebra, so it can often provide
a useful expedient for relying on the large number of results available
on the Heisenberg algebra.

Actually one can have interesting examples of pregeometric description
even based on deformed Heisenberg algebras. In particular
in Ref.~\cite{gacmaj}
(developing on results previously reported in Ref.~\cite{majidPREGEOMold})
it was noticed that one could take as starting point a two-parameter ($\rho,\hbar_{0}$)
family of commutation relations
\begin{equation}
[\hat{q},\hat{\pi}]=i\hbar_{0} (1-e^{-\frac{\hat{q}}{\rho}})
\end{equation}
with co-algebraic structure:
\begin{equation}
\Delta{\hat{q}}=\hat{q}\otimes 1 + 1\otimes \hat{q} \qquad \Delta{\hat{\pi}}=\hat{\pi} \otimes 1 + e^{-\frac{\hat{q}}{\rho}}\otimes \hat{\pi}
\end{equation}
In this setup the link between the ``pregeometric observables" $\hat{q},\hat{\pi}$ and
the  k-Minkowski coordinates emerges in the
limit\footnote{Note that in this pregeometric setup of Ref.~\cite{majidPREGEOMold}
the ``pregeometric Planck constant" $\hbar_{0}$
is in general unrelated to the physical Planck constant $\hbar$.} $\hbar_{0},\rho \to 0$, $\frac{\hbar_{0}}{\rho}\to \ell$, where one can
take $\hat{x}_{0}=\hat{\pi}$ and $\hat{x}_{1}=\hat{q}$.

Other pregeometric representations of $\kappa$-Minkowski  were developed
 (either explicitly of implicitly advocating a pregeometric viewpoint)
in Refs.~\cite{oritiPREGEOM,Agostini2007,piacitelliPREGEOM,dandreaPREGEOM}.
We shall not dwell on the details of these other pregeometric representations.
It suffices to notice that they all described  the k-Minkowski coordinates
in terms of (a few copies of) the undeformed Heisenberg algebra.
And it is also important for us to stress that these previous pregeometric
descriptions did not make room for accommodating the elements of the $\kappa$-Minkowski
differential calculus, whereas achieving a pregeometric representation
of the $\kappa$-Minkowski
differential calculus is crucial for our purposes.
Moreover, these previous studies, while providing important breakthroughs on the technical side,
left largely unaddressed the key issue for physical applications of spacetime noncommutativity:
taking as starting point our current theories where and how should we make room
for the noncommutativity of coordinates? And in which way would this noncommutativity
lead to observable effects?

\subsection{The meaning of spacetime noncommutativity}\label{subsecprevious}
Let us actually start from these last questions concerning the setting within
which spacetime noncommutativity could be both formulated and lead to observable
(at least in principle observable) consequences.
Evidently a setting of this sort cannot be based on classical mechanics, where the formalism
provides no room for noncommutativity of coordinates.
This in itself is not so alarming, since classical mechanics should anyway only emerge
as an approximate regime of a quantum mechanics, and the limiting procedure from quantum mechanics
to classical mechanics may well be such that also the noncommutativity of spacetime coordinates
is removed in the classical limit. The real problem is that even giving a formulation
of $\kappa$-Minkowski spacetime noncommutativity in a quantum-mechanics setup is not straightforward.
This is due to the fact that in $\kappa$-Minkowski the time coordinate is a noncommutative observable,
whereas in the standard formulation of quantum mechanics the time coordinate is merely an evolution
parameter (a necessarily classical evolution parameter).
Time, according to $\kappa$-Minkowski, should be an operator that does not commute with the
spatial-coordinate operators, but in the standard setup of quantum mechanics we are not in the
situation of time being described by an operator that commutes with the
spatial-coordinate operators:
in the standard setup of quantum mechanics time is not an observable at all, it just plays the role of evolution
parameter. 

We believe that it was indeed this mismatch between the nature of time in quantum mechanics
and the properties of the $\kappa$-Minkowski time coordinate that obstructed progress in formulating observable
spacetime consequences of $\kappa$-Minkowski noncommutativity.

We here propose a way to address this issue that relies on results which were not mature when $\kappa$-Minkowski
was first introduced but became increasingly solid over the last decade.
These are results~\cite{halliwellQM,gambiniportoQM,rovellireisenb}
on a covariant formulation of ordinary quantum mechanics.
In this powerful reformulation of quantum mechanics both the spatial coordinates and the time coordinate
play the same type of role. And there is no ``evolution", since
dynamics is codified in a constraint, just in the same sense
familiar for the covariant formulation of classical mechanics.
Spatial and time coordinates are well-defined operators on
a ``kinematical Hilbert space", which is just an ordinary Hilbert space of normalizable
wave functions~\cite{rovellireisenb}. And spatial and time coordinates are still well-defined operators on
the ``physical Hilbert space", obtained from the kinematical Hilbert space
 by enforcing the constraint of vanishing covariant-Hamiltonian.
 Dynamics is codified in the fact that on states of the physical Hilbert space, because
 of the implications of the constraint they satisfy, one finds relationships between the properties
 of the (partial~\cite{rovellireisenb}) observables for spatial coordinates and the properties of
 the time (partial) observable. In this way, for appropriate specification of the state on the physical Hilbert space,
 the covariant pure-constraint version of the quantum mechanics
 of free particles describes ``fuzzy worldlines" (worldlines of particles
 governed by Heisenberg uncertainty principle) just in the same sense that the covariant pure-constraint
 formulation of the classical mechanics of free particles describes sharp-classical worldlines.

So, over this last decade, the community has developed a formulation of quantum mechanics in which
both time and the spatial coordinates are operators on a Hilbert space,
which of course commute (they do not commute with their conjugate momenta,
but commute among themselves~\cite{rovellireisenb}).
Our proposal is that this is the correct starting point for formulating $\kappa$-Minkowski
noncommutativity: the commuting time and spatial-coordinate operators of the covariant formulation
of quantum mechanics should be replaced by
time and spatial-coordinate operators
governed by the $\kappa$-Minkowski noncommutativity.

Can this procedure provide us with a notion of ``geometry of $\kappa$-Minkowski"?
We believe it can. It can to the extent that a quantum spacetime can be analyzed in geometric terms.
We advocate the viewpoint
 that the kinematical Hilbert space plays a role within the covariant  formulation of quantum mechanics
 that is closely analogous to the role of Minkowski spacetime in the classical mechanics of
 special-relativistic particles.
 Within the covariant  formulation of quantum mechanics
 the kinematical Hilbert space codifies the geometry of spacetime.
 Indeed, just like Minkowski spacetime is the arena where the dynamics of relativistic
 classical particles unfolds, produced by enforcing the Hamiltonian constraint,
 the kinematical Hilbert space is the arena where the dynamics of relativistic quantum particles
 unfolds, produced by enforcing the Hamiltonian (quantum-operator) constraint.
 Minkowski spacetime on its own is not really equipped with any physically observable property:
 the observables we occasionally label as ``spacetime observables of Minkowski spacetime" truly are
 operatively defined through the experimental study of the properties of classical particles in Minkowski spacetime.
 But understanding the properties, and particularly the relativistic symmetries of Minkowski spacetime
 is an exercise of much more that mere academic interest, since the formal properties of empty Minkowski
 spacetime strongly affect then the physical properties of theories of particles in Minkowski spacetime.
 Similarly the properties of observable-operators on the
  kinematical Hilbert space of the covariant  formulation of quantum mechanics are not themselves subjectable to
  measurement,
  but they usefully characterize the spacetime arena where then the quantum dynamics of particles
  on the physical Hilbert space takes place.

So we shall here study the properties
  of the noncommuting
  coordinates of $\kappa$-Minkowski spacetime at the level of the kinematical
  Hilbert space of a  covariant  formulation of quantum mechanics. These properties of the $\kappa$-Minkowski
  coordinates will characterize $\kappa$-Minkowski spacetime as an arena for the dynamics
  of particles. We postpone the introduction of particles in $\kappa$-Minkowski ({\it i.e.} enforcing
  the Hamiltonian constraint) to a forthcoming study~\cite{usinpreparation}.
  Here we shall be satisfied with  studying the properties of the coordinates
  of 2D $\kappa$-Minkowski spacetime on the kinematical
  Hilbert space by providing a suitable ``pregeometric representation", given in terms of
  standard (undeformed) phase-space
  observables,
\begin{gather}
[\hat{\pi}_0,\hat{q}_0] = i \ , \qquad [\hat{\pi}_0 , \hat{q}_1] = 0 \nonumber \\
[\hat{\pi}_1,\hat{q}_0] = 0 \ , \qquad [\hat{\pi}_1 , \hat{q}_1] = -i \ ,
\label{pregeomPHSPACE}
\end{gather}
for the covariant formulation of 2D quantum mechanics.

\subsection{Novel pregeometric representation of $\kappa$-Minkowski coordinates, differential calculus and translations}\label{traslapresec}
For our representation of the $\kappa$-Minkowski coordinates
we view  $\hat{q}_0$ and $\hat{q}_1$ of (\ref{pregeomPHSPACE}) as operators for the pregeometric position in time
and space, indeed operators ordinarily studied~\cite{rovellireisenb}
on the kinematical Hilbert space of the covariant  formulation of quantum mechanics.
We then
describe the $\kappa$-Minkowski coordinates $\hat{x}_0,\hat{x}_1$, from (\ref{kappadefCOMMUT}),
as follows
\begin{equation}
 \hat{x}_0 = \hat{q}_0 \ , \qquad \hat{x}_1 = \hat{q}_1 e^{\ell \hat{\pi}_0 } ~,
\label{representationsXT}
\end{equation}
which indeed satisfies (\ref{kappadefCOMMUT}),

And we do find in this pregeometric description also opportunities
for describing the $\kappa$-Minkowski differential calculus and
the $\kappa$-Poincar\'e translation generators.
For the translation generators
by posing
\begin{gather}
 P_{0}\triangleright f(\hat{x}_0 , \hat{x}_1) \longleftrightarrow
  [\hat{\pi}_{0},f(\hat{q}_0 ,  \hat{q}_1 e^{\ell \hat{\pi}_0 })] \ , \nonumber \\
  P_1 \triangleright f(\hat{x}_0 , \hat{x}_1) \longleftrightarrow
   e^{- \ell \hat{\pi}_{0}}[\hat{\pi}_{1},f(\hat{q}_0 ,  \hat{q}_1 e^{\ell \hat{\pi}_0 })] ~,
\label{representationP}
\end{gather}
one does reproduce all the properties of $\kappa$-Poincar\'e translation generators,
here summarized in Sec.~\ref{seckappafacts}.

And most crucially we also notice that the properties of the elements $\hat{a}_\mu^\ell$
of the differential calculus
given in Eq.~(\ref{parametersCOMM}) can be reproduced
by combining ordinary (numerical) parameters $a_\mu$ and the observable $\hat{\pi}_0$:
\begin{equation}
\hat{a}_0^\ell = a_0 \ , \qquad \hat{a}_1^\ell = a_1 e^{\ell \hat{\pi}_0 } ~.
\label{representationPARAM}
\end{equation}

\section{Boosts and a fully pregeometric picture}
We have so far ignored boosts, though they are also known~\cite{majrue} to be strongly
affected by $\kappa$-Minkowski noncommutativity.
The main objectives of the analysis we are here reporting concern
translation transformations, which, as shown in the next section, when formulated according
to our proposals and pregeometric formulation, shed light on several grey areas of our
previous understanding of $\kappa$-Minkowski.
But before we get to that let us pause in this section for introducing our pregeometric
description of boosts in $\kappa$-Minkowski, not only for showing the completeness of our
pregeometric representation, but also
for completing the characterization of the kinematical
Hilbert space on which this manuscript focuses.

We have already implicitly specified that the states of our kinematical Hilbert space for $\kappa$-Minkowski
will admit representation (in the ``pregeometric momentum-space representation") as square-integrable
functions of variables $\hat{\pi}_0$ and $\hat{\pi}_1$. But the prescription of square-integrability is meaningful only
once a measure
on this kinematical Hilbert space is introduced. Understanding the properties
of boosts in $\kappa$-Minkowski, as formulated in our pregeometric picture,
will allow us to specify this measure and we shall see that in that respect
our kinematical Hilbert space is not exactly the same as the one
 (see, {\it e.g.}, Ref.~\cite{rovellireisenb})
 of the covariant formulation
of quantum mechanics.

Essentially the task we must accomplish is providing a pregeometric description
 of the boost sector of the $\kappa$-Poincar\'e Hopf algebra.
Working again consistently with the choice of conventions introduced in  Ref.~\cite{majrue},
which we adopt throughout,
in our 2D $\kappa$-Minkowski spacetime boost generators should satisfy the following
properties of commutation with translation generators and of coproduct:
\begin{eqnarray}
  && -i [N,P_0] \triangleright f(\hat{x}) \equiv P_1 \triangleright f(\hat{x}) \ ,
  \label{boostcommut0}\\
  && -i [N,P_1] \triangleright f(\hat{x})\equiv \left( \frac{1-e^{-2\ell P_0}}{2\ell} - \frac{\ell}{2} P_1^2  \right) \triangleright f(\hat{x}) \ ,
\label{boostcommutJ}
\end{eqnarray}
\begin{equation}
\Delta N = N \otimes \mathbbm{1} + e^{-\ell P_{0}} \otimes N \ ,
\label{coproductBOOST}
\end{equation}
Notice that in the 2D $\kappa$-Minkowski the coproduct of boost generators has the same
form as the coproduct of translation generators (here shown in Sec.~\ref{seckappafacts}).
This is a peculiarity of the 2D case which simplifies the description of boost transformations.
Generalizing our results for the pregeometry from our 2D case to a 4D $\kappa$-Minkowski
for what concerns translation transformations is completely elementary.
For boosts the 4D generalization is also conceptually straightforward but
technically requires the added structure
of the specific properties of $\kappa$-Poincar\'e boosts in the 4D case, where the
coproduct of boost generators no longer has the same structure of the coproduct of translation generators.

This fact that 2D $\kappa$-Poincar\'e boost generators have the same coproduct as
2D $\kappa$-Poincar\'e  translation generators immediately leads us to also specify
the properties of boost transformation parameters. In fact, as observed for example
in Ref.~\cite{nopure}, the noncommutativity properties of transformation parameters
are directly linked to the coproduct properties of the generators of the transformations.
Also for boost transformations we can give a formulation analogous to
the one of Eq.~(\ref{translationGENERATOR}), with the action of  boosts taking the form
\begin{equation}
B=\mathbbm{1} + {\bf d}_N \ , \qquad {\bf d}_N = i \hat{\xi}^\ell N \ ,
\label{boostGENERATOR}
\end{equation}
and noncommutative boost-transformation parameter such that~\cite{nopure}
\begin{equation}
\left[ \hat{\xi}^\ell , \hat{x}_0 \right] = i\ell \hat{\xi}^\ell ~ ,~~~~\left[ \hat{\xi}^\ell , \hat{x}_1\right] = 0 \ .
\label{BOOSTparametersCOMM}
\end{equation}

Our task then is to provide a pregeometric representation of
the boost generator $N$ and of the noncommutative boost-transformation parameter $\hat{\xi}^\ell$
reflecting the properties  (\ref{boostcommutJ}), (\ref{coproductBOOST}),
(\ref{BOOSTparametersCOMM}).
We find that this is indeed possible. The pregeometric description of
the noncommutative boost-transformation parameter $\hat{\xi}^\ell$ is given
in terms of an ordinary (numeric, commutative) boost parameter $\xi$ and
the $\hat{\pi}_0$ observable
\begin{equation}
\hat{\xi}^\ell = \xi e^{\ell \hat{\pi}_0} ~.
\label{representationPARAMboost}
\end{equation}
For the boost generator we find the pregeometric prescription
\begin{equation}
  N\triangleright f(\hat{x}) \equiv e^{-\ell \hat{\pi}_{0}}[\hat{\eta},f(\hat{x})]
\label{boostdef}
\end{equation}
with
\begin{equation}
\hat{\eta} \equiv \left(\frac{e^{2\ell \hat{\pi}_{0}}-1}{2\ell} + \frac{\ell}{2}\hat{\pi}_{1}^{2}\right)\hat{q}_{1} - \hat{\pi}_{1}\hat{q}_{0}
\label{etadef}
\end{equation}
These pregeometric representations provide the basis for studying  boost
transformations in $\kappa$-Minkowski. And it is valuable to notice 
that from (\ref{representationPARAMboost}) and (\ref{boostdef}) it follows that
 the action (\ref{boostGENERATOR}) can be expressed in terms of the operator $\hat{\eta}$ as an ordinary (adjoint) action by commutator, which in particular can be exponentiated as usual
\begin{equation}
B\triangleright\hat{O} \rightarrow {\cal{B}}^\dagger \hat{O} {\cal{B}} = e^{i\xi\hat{\eta}^\dagger} \hat{O} e^{-i\xi\hat{\eta}} \ .
\label{BoostOPERATOR}
\end{equation}

At this point we have exhibited the full strength
of our pregeometric description: whereas previous pregeometric descriptions only accommodated the $\kappa$-Minkowski
coordinates (plus, in some cases, some $\kappa$-Poincar\'e generators)
we gave a pregeometric description of all
the most used tools of the literature on $\kappa$-Minkowski, including the differential calculus (which also play the role of noncommutative translation parameters),
the translations generators,
the noncommutative boost parameter and the boost generator.

And we are now well equipped for returning to the issue highlighted at the beginning
of this subsection, concerning the specification of the measure on
our kinematical Hilbert
space. We shall characterize our scalar products in momentum space, as
\begin{equation}
\langle \hat{O} \rangle = \langle \psi | \hat{O} | \psi \rangle = \int {\cal D}(\pi_\mu) \psi^\star(\pi_\mu) O(\pi_\mu) \psi(\pi_\mu) ~,
\end{equation}
and in order to get a boost invariant scalar product, we want the measure ${\cal D}(\pi_\mu)$ to be invariant under the action of boosts (\ref{boostGENERATOR}). From the definitions
(\ref{boostGENERATOR}),(\ref{representationPARAMboost}),(\ref{boostdef}),(\ref{etadef}), together with (\ref{pregeomPHSPACE}), one easily finds that under the action of
boosts (\ref{boostGENERATOR})
\begin{gather}
\pi_{0}'=\pi_{0}-\xi\pi_{1}\ ,\nonumber \\
\pi_{1}'=\pi_{1}-\xi\left(\frac{e^{2\ell\pi_{0}}-1}{2\ell}+\frac{\ell}{2}\pi_{1}^{2}\right)\ .
\label{boostACTIONmomenta}
\end{gather}
Guided by the criterion that the measure ${\cal D}(\pi_\mu)$ should
 be invariant under these transformations we are led to adopt
\begin{equation}
{\cal D}(\pi_\mu) = d\pi_0 d\pi_1 e^{-\ell \pi_0}\ .
\label{measureDEFORMED}
\end{equation}
One easily sees that this  deformed measure (\ref{measureDEFORMED})
also ensures that $\hat{\eta}$ is hermitian, so that
the boost operator ${\cal{B}}$, introduced in (\ref{BoostOPERATOR}), is unitary
and preserves the scalar product:
\begin{equation}
\langle \psi' | \psi' \rangle = \langle \psi |e^{i \xi \hat{\eta}} e^{-i \xi \hat{\eta}} | \psi \rangle = \langle \psi | \psi \rangle \ .
\end{equation}

While, as mentioned, we postpone to a forthcoming study~\cite{usinpreparation}
the introduction of physical/on-shell particles in our $\kappa$-Minkowski spacetime,
let us pose briefly for observing that the properties of boosts also strongly
characterize the form of the on-shell condition, which in turn (through an appropriate ``Hamiltonian constraint"~\cite{rovellireisenb}) governs the relationship
between the kinematical Hilbert space and the physical Hilbert space.
On the basis of the properties derived above one easily finds that
the demand of invariance under boosts leads to adopting the
following ``deformed d'Alembertian operator"
\begin{equation}
 \square_{\ell}= \left(\frac{2}{\ell}\right)^{2}\sinh^{2}\left(\frac{\ell \hat{\pi}_{0}}{2}\right) - e^{-\ell \hat{\pi}_{0}}\hat{\pi}_{1}^{2} ~.
\label{dalembertell}
\end{equation}

We are postponing these issues for physical particles and the physical
Hilbert space, since the main objectives of this manuscript can be pursued by studying
the properties of the $\kappa$-Minkowski coordinates
on our kinematical Hilbert space. These are the properties which, in the sense discussed
in the previous section, we view as the way to describe the geometry
of empty $\kappa$-Minkowski space (to the extent that this still can make sense
when speaking of a nonclassical geometry).

For these tasks which are the main focus of the remainder of this manuscript
it is useful to assess the implications of the
integration measure ${\cal D}(\pi_\mu)$ of (\ref{measureDEFORMED})
for the properties of
the  $\kappa$-Minkowski coordinates.
One easily sees that the spatial coordinate  $\hat{x}_1$ is a hermitian
operator on our kinematical Hilbert space equipped with the integration
measure ${\cal D}(\pi_\mu)$, since, by the momentum space representation $\hat{q}_1 \equiv i \partial_{\pi_1} $, it follows that
$$
\hat{q}_1 e^{-\ell \pi_0} = e^{-\ell \pi_0} \hat{q}_1 \ .
$$

For the operator  $\hat{q}_0$ one easily sees
that it is not hermitian but it misses being hermitian by a constant
term of order $\ell$
$$\hat{q}_0^\dagger = \hat{q}_0 + i\ell ~.$$
Indeed, since $\hat{q}_0 \equiv - i \partial_{\pi_0} $, one finds that
$$
\hat{q}_0 e^{-\ell \pi_0} = e^{-\ell \pi_0} (\hat{q}_0 + i\ell)\ .
$$

We shall not be too concerned about this peculiar lack of hermitianity of $\hat{q}_0$.
One could easily obtain from $\hat{q}_0$ a hermitian operator that can
serve the purpose of $\kappa$-Minkowski time coordinate\footnote{Note that $\hat{x}_0^*$ is a good choice
of $\kappa$-Minkowski time coordinate, since $[\hat{x}_1,\hat{x}_0^*] = i \ell \hat{x}_1$.
 And one also easily verifies
 that $\kappa$-Minkowski described by $\hat{x}_1,\hat{x}_0^*$ has good properties
 under boosts, $[B \triangleright \hat{x}_1, B \triangleright \hat{x}_0^*] = i \ell B \triangleright \hat{x}_1$
 (or $N\triangleright [\hat{x}_1,\hat{x}_0^*] = i\ell N \triangleright \hat{x}_1 $),
 and under
 translations, $[T \triangleright \hat{x}_1, T \triangleright \hat{x}_0^*] = i \ell T \triangleright \hat{x}_1$
 (or $P_\mu \triangleright [\hat{x}_1,\hat{x}_0^*] = i\ell P_\mu \triangleright \hat{x}_1 $).},
such as $\hat{x}^*_0 \equiv \hat{q}_0 - i \ell/2$.
But we feel we can still take $\hat{x}_0 = \hat{q}_0$ since the properties
of $\hat{x}_0$ on our kinematical Hilbert space are not truly observable: they merely
provide a way for characterizing the abstract notion of the geometry
of empty $\kappa$-Minkowski spacetime.
As we shall show in Ref.~\cite{usinpreparation}
the physical properties of $\kappa$-Minkowski spacetime, the ones affecting
the analysis of the physical Hilbert space, will have to be formulated
in terms of operators that commute with the Hamiltonian constraint, written in terms
of (\ref{dalembertell}), and
the $\kappa$-Minkowski time coordinate is not one such observable.
Moreover, when we are interested in the $\kappa$-Minkowski time coordinate
as a partial observable~\cite{rovellireisenb} on the physical Hilbert space
we shall inevitably find that the most meaningful
features are to be phrased in terms of differences among values of
this operator (reflecting the evident fact that the physical content of the notion of ``time" all
 resides in time differences/intervals),
so that the choice between  $\hat{x}_0 = \hat{q}_0$
and $\hat{x}^*_0 \equiv \hat{q}_0 - i \ell/2$
is intangible.

\section{Fuzzy points, translation transformations and relative locality}
In the previous sections we introduced  a novel pregeometric description
of all the ingredients
needed for describing ``points" in $\kappa$-Minkowski and for examining these fuzzy points
from the perspectives of pairs of distant observers in relative rest, observers
connected by a pure translation. In this section we shall show that $\kappa$-Minkowski,
contrary to what might appear when looking naively at its commutation relations,
affords us a fully relativistic description of distant observers, and
provides the first ever example of relative locality in a quantum spacetime.

\subsection{Fuzzy points}

First we need to give a description of points in $\kappa$-Minkowski. We of course
expect that they should not be the sort of sharp points available in a classical geometry.
Evidently within our pregeometric description a point will be identified with
a state in the pregeometric Hilbert space that gives rather well determined values
to $\hat{x}_0$ and $\hat{x}_1$. It is indeed easily seen that
no state in the pregeometric Hilbert space
 gives absolutely sharp values to $\hat{x}_0$ and $\hat{x}_1$: in light
 of $\hat{x}_0 = \hat{q}_0$, $\hat{x}_1 = \hat{q}_1 e^{\ell \hat{\pi}_0 }$ a sharply specified $\hat{x}_0$ requires an eigenstate
of $\hat{q}_0$ but on such eigenstates of $\hat{q}_0$ one has that $\hat{\pi}_0$ is infinitely fuzzy
($\delta \pi_0 \sim \infty$) which in turn implies that $\hat{x}_1 = \hat{q}_1 e^{\ell \hat{\pi}_0 }$
cannot be sharp. So all points in $\kappa$-Minkowski must be fuzzy\footnote{We shall pay little
attention to the fact that actually there is an exception to this ``fuzziness theorem":
the interested reader can easily verify that the origin of the
observer, $x_0=x_1=0$, can be sharp.
This can be straightforwardly added as a limiting case for the discussion we offer in the following,
and in particular one finds that even a point that is absolutely sharp in the origin
of one observer is described by a distant observer as a fuzzy point.}.

A class of pregeometric states which is well suited for exploring the properties
of $\kappa$-Minkowski fuzziness is the one of gaussian states on our pregeometric Hilbert space.
 We adopt a ``pregeometric-momentum-space description" of these gaussian states,
denoted by $\Psi_{\overline{\pi}_\mu , \sigma_\mu, \overline{q}_\mu} (\pi_\mu)$
so they are given in terms of functions of the variables $\pi_\mu$ parametrized
by $\overline{\pi}_\mu$ , $\sigma_\mu$, and $\overline{q}_\mu$:
\begin{equation}
\Psi_{\overline{\pi}_\mu , \sigma_\mu, \overline{q}_\mu} (\pi_\mu)
\!=\!\! N \! e^{- \frac{\left(\pi_0 - \overline{\pi}_0\right)^2}{4\sigma_0^2} - \frac{\left(\pi_1 - \overline{\pi}_1\right)^2}{4\sigma_1^2}} \!\! e^{ i \pi_0 \bar{q}_0 - i \pi_1 \bar{q}_1},
\label{state}
\end{equation}
where $N$ is a normalization constant.
Essentially $\overline{\pi}_0,\overline{\pi}_1$ have the role of expected values for
the pregeometric momenta $\hat{\pi}_0 , \hat{\pi}_1$, whereas
$\sigma_0,\sigma_1$ characterize the uncertainties for $\hat{\pi}_0 , \hat{\pi}_1$.
Moreover, as we shall see, $\overline{q}_0,\overline{q}_1$
determine the expected values for
the pregeometric position coordinates $\hat{q}_0 , \hat{q}_1$.

Of course, our main focus of attention will be on establishing how the $\kappa$-Minkowski
scale $\ell$ affects the results. This is going to be our indicator
of the difference between classical Minwkoswki spacetime and $\kappa$-Minkowski.
We start by noting down how the scale $\ell$ intervenes in the normalization
factor $N$. By imposing $\langle \Psi | \Psi \rangle = 1$, and taking into account
the integration measure (\ref{measureDEFORMED}), one easily finds that
\begin{equation}
N^2 = \frac{e^{\ell\bar{\pi}_{0}}e^{-\frac{\ell^{2}\sigma_{0}^{2}}{2}}}{2\pi \sigma_{0}\sigma_{1}} ~.
\label{normalization}
\end{equation}

We characterize the properties of  points of $\kappa$-Minkowski spacetime
by evaluating in our gaussian pregeometric states (\ref{state})
the mean values and uncertainties of the operators $\hat{x}_0$, $\hat{x}_1$
as $\bar{x}_0=\langle \hat{q}_0\rangle$ and $\bar{x}_1=\langle \hat{q}_1 e^{\ell \hat{\pi}_0 } \rangle$, while for the uncertainties $\delta \hat{x}_0$,$\delta \hat{x}_1$
we can resort to
 $\delta \hat{x}_0= \sqrt{\langle \hat{q}_0^2\rangle - \bar{x}_0^2}$
  and $\delta \hat{x}_1
  = \sqrt{\langle \left(\hat{q}_1 e^{\ell \hat{\pi}_0 }\right)^2\rangle - \bar{x}_1^2}$.

Unsurprisingly the  $\kappa$-Minkowski
scale $\ell$ turns out to play a particularly significant role in the properties
of the coordinate  $\hat{x}_1$, for which we find
\begin{gather}
\langle \hat{x}_1 \rangle = \langle \hat{q}_1 \rangle \left\langle e^{ \ell \hat{\pi}_0} \right\rangle = \overline{q}_1 e^{\ell\bar{\pi}_{0}} e^{-\frac{\ell^{2}\sigma_{0}^{2}}{2}} \ , \nonumber \\
\delta \hat{x}_1 = e^{\ell\bar{\pi}_{0}} \left[ \frac{1}{4\sigma_1^2} + \overline{q}_1^2\left( 1 - e^{-\ell^2 \sigma_0^2}\right) \right]^{1/2}  ~ .
\label{meanX}
\end{gather}
Instead for the $\kappa$-Minkowski time coordinate $\hat{x}_0$
there is no $\ell$-deformation, with the exception of the constant imaginary contribution
of order $\ell$ which should be expected on the basis of the remarks
at the end of the previous section  (and to which we attach little significance,
for reasons also stressed at the end of the previous section):
\begin{equation}
\langle \hat{x}_0 \rangle = \overline{q}_0 - i \frac{\ell}{2} \ ,\qquad \delta \hat{x}_0 = \frac{1}{2\sigma_0} ~,
\label{meanT}
\end{equation}
We notice already at this stage that for fixed values
of  $\overline{q}_0,\overline{\pi}_0,\sigma_0,\sigma_1$
one finds larger fuzziness of $\hat{x}_1$ at large values of $\overline{q}_1$,
because of the contribution to $\delta \hat{x}_1$ by the term with
$\overline{q}_1^2$
%$\overline{q}_1^2\left( 1 - e^{-\left(\ell\sigma_{\pi_0}\right)^2}$
in (\ref{meanX}).
It is however of very limited interest to compare different fuzzy points
in $\kappa$-Minkowski: at any distance from the origin we can anyway get points
as fuzzy as we might desire. The key feature we need to uncover concerns
how the same point is seen by observers close to it  and by distant observers.

\subsection{Translations and Relative Locality}\label{translations}
The pregeometric description given in Subsec.~\ref{traslapresec} already
provides us all that is needed for implementing a translation transformation
on one of our fuzzy $\kappa$-Minkowski points.
Evidently the action of our translations will be of the form
$T=\mathbbm{1} + {\bf d}_P$, with ${\bf d}_P= - i a^\ell_\mu P^\mu$,
and we established in Subsec.~\ref{traslapresec} that
$P_{0}\triangleright f(\hat{x}) \equiv [\hat{\pi}_{0},f(\hat{x})]$
and $P_1 \triangleright f(\hat{x}) \equiv e^{- \ell \hat{\pi}_{0}}[\hat{\pi}_{1},f(\hat{x})]$
whereas
$\hat{a}_0^\ell = a_0$, $\hat{a}_1^\ell = a_1 e^{\ell \hat{\pi}_0 }$.
In particular, our construction  exposed
the simplicity of the differential operator ${\bf d}_P$, which in previous
works on $\kappa$-Minkowski remained hidden behind the rather virulent properties
of the generators $P_\mu$ and of the elements $\hat{a}_\mu^\ell$
of the noncommutative differential calculus:
within our pregeometric description one has that
\begin{equation}
{\bf d}_P \triangleright f(\hat{x}_0 , \hat{x}_1) \longleftrightarrow
 -i a^\mu \left[\hat{\pi}_\mu ,f(\hat{q}_0 ,  \hat{q}_1 e^{\ell \hat{\pi}_0 })\right]~,
\label{representationD}
\end{equation}
so this action involves only familiar commutative transformation parameters $a_\mu$
and standard translations (acting by commutator) at the pregeometric level.

This allows us to implement translation transformations straightforwardly. We find
\begin{gather}
T\triangleright \hat{x}_0 = \hat{x}_0 - \hat{a}_0^\ell = \hat{q}_0 - a_0\ , \nonumber \\
T\triangleright \hat{x}_1 = \hat{x}_1 - \hat{a}_1^\ell = e^{\ell \hat{\pi}_0}\left(  \hat{q}_1 - a_1 \right)\ .
\end{gather}

We can now evaluate the mean values and uncertainties of $T \triangleright \hat{x}_\mu$ on the gaussian state (\ref{state}), to find\footnote{Of course the same results for mean values and uncertainties
of $\kappa$-Minkowski coordinates can be obtained by acting with $T$ on the pregeometric state
and evaluating $\hat{x}_\mu$ and $\delta \hat{x}_\mu$ in the state thereby obtained. The equivalent alternative
we follow, by acting with $T$ on $\hat{x}_\mu$ and evaluating the mean value and the
uncertainty of  $T\triangleright \hat{x}_\mu$ in the original state just allows the derivation to proceed
a bit more speedly. }
\begin{gather}
\langle T\triangleright \hat{x}_0 \rangle = \overline{q}_0 - a_0 -i\frac{\ell}{2}
\label{meanTTa}\\
 \delta \left( T\triangleright \hat{x}_0  \right)= \frac{1}{2\sigma_0} \ ,
\label{meanTTb}
\end{gather}
and
\begin{gather}
\langle T\triangleright \hat{x}_1 \rangle = \left( \overline{q}_1 - a_1 \right)e^{\ell \overline{\pi}_0} e^{-\frac{\ell^{2}\sigma_{0}^{2}}{2}} \label{meanTXa}\\
\delta \! \left( T\triangleright \hat{x}_1  \right) \!=\! e^{\ell \overline{\pi}_0} \!\left[ \frac{1}{4\sigma_1^2} \!+\! \left(\overline{q}_1 \!-\! a_1 \right)^2 \!\left( 1 \!-\! e^{-\ell^2 \sigma_0^2} \right) \right]^{\! 1/2} \!\!\!\!.
\label{meanTXb}
\end{gather}

The interpretation here of course is such that the $\hat{x}_\mu$ are operators
characterizing the distance of a given (fuzzy)
point from the frame origin of
some observer Alice, and then  $T\triangleright \hat{x}_\mu$ are the operators
that characterize the distance
of that point from the frame origin of an observer Bob,
purely translated with respect to Alice.
 And accordingly
 one can deduce the relation between the mean values and uncertainties in positions
among two distant observers in relative rest by comparing (\ref{meanT}) to (\ref{meanTTa})-(\ref{meanTTb}) and
 comparing  (\ref{meanX}) to (\ref{meanTXa})-(\ref{meanTXb}).

The main message is contained in our Fig.~1. There we show two fuzzy points
in $\kappa$-Minkowski as described by two distant observers.
One of the points is near observer Alice, while the other one is near
observer Bob, purely translated with respect to Alice.
What is shown in figure is indeed obtained by comparing
(\ref{meanT}) to (\ref{meanTTa})-(\ref{meanTTb}) and
 comparing  (\ref{meanX}) to (\ref{meanTXa})-(\ref{meanTXb}).
 There are two main features:\\
 (i) the same point appears to be more fuzzy to a distant observer than to a nearby observer, and\\
(ii) the point at Alice  is not described as being at Alice in the
coordinatization of spacetime of observer Bob, and {\it vice versa}
the point at Bob is not described as being at Bob in the
coordinatization of spacetime of observer Alice.\\
The second feature, (ii), is essentially already known from previous studies
of relative locality in the classical limit~\cite{bob,principle}: one
can have consistently relativistic theories where pairs of points found to be coincident
by a nearby observer (or, as in the case here considered, a point found to coincide with
the origin of that observer) are instead described as noncoincident if one uses the inferences
about those points by a distant observer. Feature (i) is here established for the first
time in the literature. It is a feature of relative locality for the fuzziness of points
in a quantum spacetime. And, as shown in figure, it is also a fully relativistic effect: on the basis
of the content of Fig.~1 there is no way to distinguish between Alice and Bob.
Alice attributes to the point at Bob more fuzziness than observed by Bob, and also
Bob attributes to the point at Alice more fuzziness than observed by Alice.
This unveils the nature of $\kappa$-Minkowski as a fully relativistic spacetime.
Without our more powerful characterization of
$\kappa$-Minkowski one could (see {\it e.g.} Ref.~\cite{oritiPREGEOM})
think that ``$\kappa$-Minkowski has a special point, a sort of center, and the
origin of the $\kappa$-Minkowski preferred frame should be made coincide with that
special point". Instead we can now clearly see that no point is special and no observer
is special/preferred in $\kappa$-Minkowski: all observers in $\kappa$-Minkowski
have the property that they perceive their origin as the point of lowest fuzziness
and attribute to distant points fuzziness proportional to the distance (but then
an observer
located at one of those distant points will again describe that point as the one of minimal
fuzziness).

 \newpage

\begin{figure}[h!]
\includegraphics[width=\columnwidth]{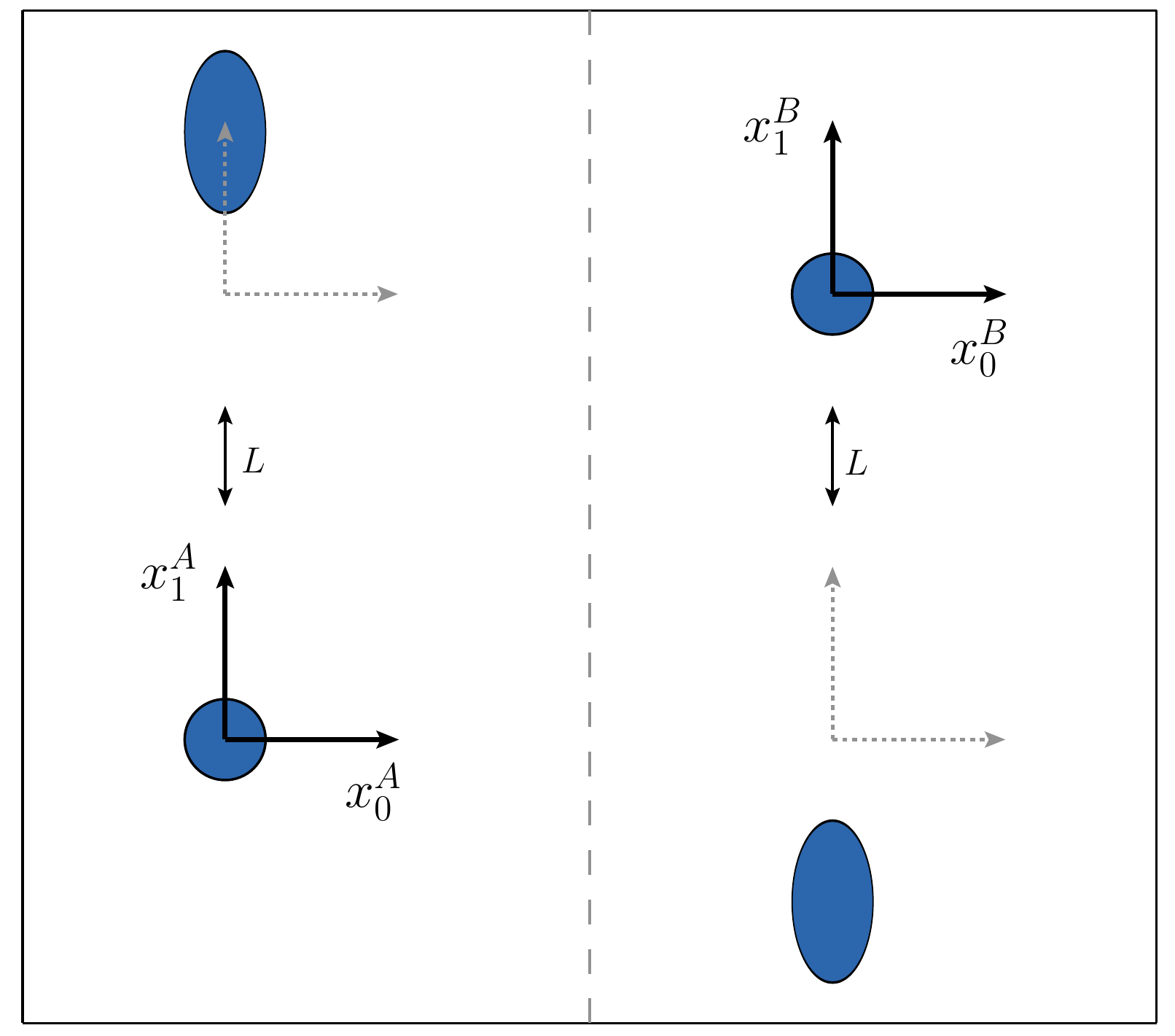}
\caption{We illustrate the features of relative locality we uncovered for the $\kappa$-Minkowski quantum spacetime
by considering the case of two distant observers, Alice and Bob, in relative rest (with synchronized clocks).
In figure we have only two points in $\kappa$-Minkowski, each described by a gaussian state in our Hilbert space.
One of the points is at Alice (centered in the spacetime origin of Alice's coordinatization)
while the other point is at Bob. The left panel reflects Alice's description of the two points, which in particular
attributes to the distant point at Bob larger fuzziness than Bob observes (right panel). And in Alice's coordinatization the distant point is not exactly at Bob. Bob's description (right panel)
of the two points is specular,
in the appropriately relativistic fashion, to the one of Alice. The magnitude of effects shown would require
the distance $L$ to be much bigger than drawable. And for definiteness in figure we
assumed $\bar{\pi}_0 \simeq 2 \sigma_0$ and $ \sigma_1 \simeq  \sigma_0$.}
\end{figure}

\vskip 2.8cm

As for other aspects of relative locality
one gains some insight~\cite{bob,principle}
by comparing the case of the implications for relative locality of the
introduction of the relativistic invariant
(inverse-)momentum scale $\ell$
to the familiar case of the implications for relative simultaneity
of the introduction of the relativistic invariant speed (-of-light) scale $c$.
We show in Fig.~2
an aspect of relative simultaneity that can be viewed from a perspective that
is somewhat analogous to the features of relative locality here highlighted
in the previous Fig.~1. It is a situation such that Alice and Bob are once again
in relativistically specular conditions (in Fig.~1 the specularity concerned distances,
since it is primarily translation transformations that need to be deformed to accommodate
relative locality, whereas in Fig.~2 the specularity concerns relative speeds,
since boost transformations are to be deformed, with respect to the Galilean absolute-simultaneity
case, in order to accommodate relative simultaneity). In Fig.~2 Alice and Bob make stipulations
specified as rest frame properties (e.g. the laser light wavelength in the rest frame of the laser)
and ordinary special relativity, with its relativity of simultaneity, produces several features
which would appear to be paradoxical to a Galilean scientist.

\begin{figure}[h!]
\includegraphics[scale=0.8]{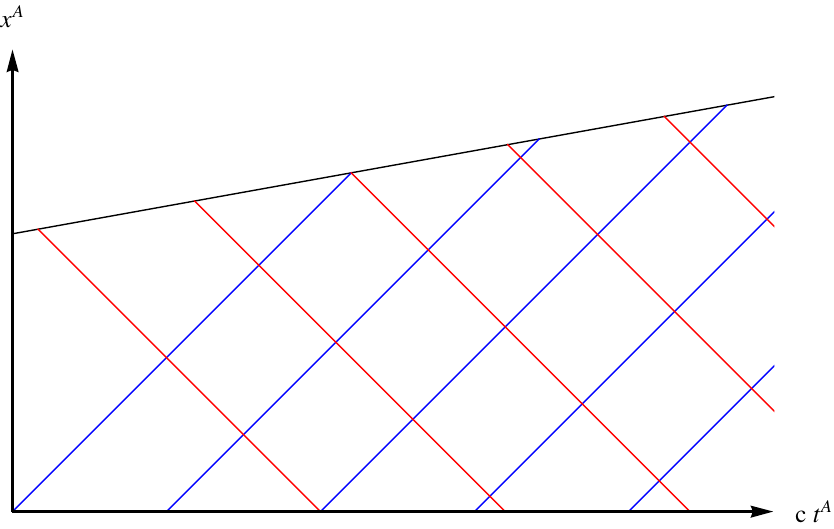}
\includegraphics[scale=0.8]{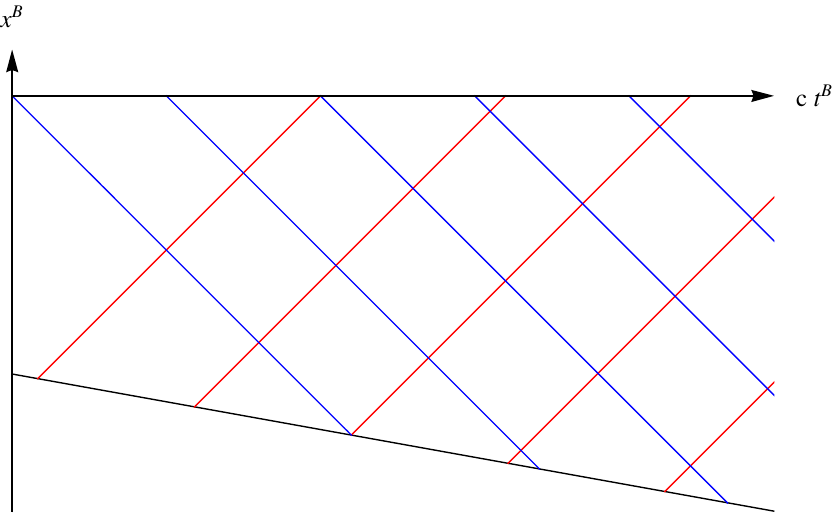}
\caption{Here Alice (coordinatization shown in top panel)
and Bob (coordinatization shown in bottom panel)
are evidently in relative motion with constant speed. The worldlines
of massless particles are described (in blue and red) here assuming the validity
of ordinary special relativity. Alice and Bob have stipulated a procedure of clock
synchronization and agreed to build  emitters  of  blue photons (blue
according to observers at rest with
respect to the emitter). They also agreed to emit such blue photons in a regular
sequence, with equal time spacing $\Delta t^*$. We arranged the starting time of each sequence of emissions
so that there would be two cases of a detection  coinciding with an emission. These coincidences of events
are of course manifest in both coordinatizations (special relativity is absolutely local).
But relative simultaneity is directly or indirectly
responsible for several features
that would appear to be paradoxical to a Galilean observer (observer assuming absolute simultaneity).
In particular, while they stipulated to build blue-photon emitters they detect red photons,
and while the emissions are time-spaced by $\Delta t^*$
the detections are separated by a time greater than $\Delta t^*$.}
\end{figure}

\newpage

\section{Outlook}
We have here set the stage for studies of the $\kappa$-Minkowski noncommutative spacetime to finally
be focused on the implications as a description of ``quantum spacetime".
One of the most robust indications we have about the Planck-scale realm is indeed that spacetime
should be ``quantized", described by a nonclassical geometry. And we need the guidance of some
models in order to sharpen our understanding of this new notion and possibly for devising
dedicated phenomenological programmes. We feel $\kappa$-Minkowski should now prominently be listed
among the key illustrative examples of what spacetime quantization might bring about.

On the technical side the main lesson we draw from our analysis is that ``pregeometric" representations
intended for the study of the $\kappa$-Minkowski spacetime are of little use if the picture does
not make room for a representation of the differential calculus, a first example of which
was given here.
We expect that this point should apply, suitably adapted, also to other noncommutative spacetimes.

Among the features we uncovered of particular conceptual interest is our description
of $\kappa$-Minkowski as a fully relativistic spacetime,
well suited as a case study for the concept of DSR-deformed relativistic
symmetries~\cite{gacdsr1a,jurekDSRnew,leeDSRprd,gacdsrrev2010}.
The possibility that theories in $\kappa$-Minkowski might be examples of DSR theories
had been suggested in several previous studies (see, {\it e.g.}, Refs.~\cite{gacdsr1a,jurekDSRnew,gacdsrrev2010}),
but only on the basis of semiheuristic arguments centered on the expected properties
of a momentum space dual to $\kappa$-Minkowski. Those suggestions remained {\it sub judice} because
of several missing ingredients~\cite{gacdsrrev2010}, within available $\kappa$-Minkowski results,
of a comprehensively relativistic picture, and among these ``open issues" for a relativistic
description of $\kappa$-Minkowski of particular concern were the features connected with
the apparently special role of the origin~\cite{oritiPREGEOM}.
Our analysis contributes significantly toward establishing $\kappa$-Minkowski as a DSR-relativistic
spacetime, and crucial for this was recognizing that no point is special in $\kappa$-Minkowski:
each point in $\kappa$-Minkowski can serve equally well as the origin of some observer, and then
for that observer it acquires an unsurprisingly privileged role.
Awareness of the possibility of a relativity of spacetime locality was crucial for uncovering
these features. And this could encourage other studies tailored to exploit awareness
of the possibility of relative locality for improved understanding of some quantum spacetimes,
 whereas before the analysis here reported relative locality had only been considered
 within theories of classical point particles.

\section*{Acknowledgements}
We gratefully acknowledge conversations with Daniele Oriti and Carlo Rovelli.
This work was
supported in part by a grant from the John Templeton Foundation.

\end{document}